\documentclass[prl,twocolumn,aps]{revtex4}
\usepackage{amsmath}
\usepackage{amssymb}
\usepackage{graphicx}

\begin{document}
\title{Aharonov-Bohm effect mediated by massive photons
}
\author{Kicheon Kang}
\email{kicheon.kang@gmail.com} 
\affiliation{Department of Physics, Chonnam National University, Gwangju 61186, 
 Republic of Korea}

\begin{abstract}
Virtual photons play an essential role in the locally realistic description
of the Aharonov-Bohm interference. We show that the
effect of virtual photons in the interferometer 
is manifested by a change in their spectrum.
In particular, when a vacuum is confined between two
ideal conducting plates, the photons obey the two-dimensional Proca equation,
the wave equation with finite effective mass. This results in a short-range 
interaction between a test charge and a magnetic flux, and hence the
Aharonov-Bohm effect is reduced exponentially at a large distance 
between the two bodies. On the other hand, a semiclassical description is
also possible, and this raises the interesting question of how to prove the
physical reality of virtual photons.
\end{abstract}

\maketitle

{\em Introduction-}.
The modern quantum field theory view of fundamental interactions 
is that particles of matter
do not interact directly with each other,
but exchange virtual particles (gauge bosons), which are the mediators of
the interactions.
In the case of electromagnetic interaction, the mediators are photons,
and the long-range nature of the  
electromagnetic interaction arises from the massless spectrum of the photons.
This picture of electrodynamic interaction mediated by virtual photons 
should also apply to the case 
where the mutual force between two distant bodies disappears.
A particularly interesting example is the Aharonov-Bohm~(AB) 
effect~\cite{aharonov59,ehrenberg49} 
which is often cited as part of the quantum weirdness, specifically as
evidence for nonlocal action~\cite{aharonov16} between the two bodies. 
There has been a recent revival of debates about the nature of 
locality in the AB effect~\cite{aharonov16,vaidman12,kang13,kang15,kang17}.
Quantum electrodynamics (QED) provides a microscopic picture
that can solve the locality problem 
of the AB effect~\cite{marletto20,saldanha21,kang22}.
In QED,
the interaction between a charged particle and a distant magnetic flux
is not associated with an instantaneous and nonlocal action, but results from 
the exchange of virtual photons~\cite{santos99,lee04}.
In other words, QED gets rid of the ``action at a distance" inherent in
the conventional semiclassical approach to the AB effect.
Moreover, this is not just a matter of interpretation, but can be verified
by measuring the gauge-invariant local phase shift~\cite{kang17,kang22}.

While this suggests the crucial role of virtual photons in understanding
the AB effect, observing its manifestation is another matter. 
This could be achieved by modifying the spectrum of the photon, since the latter
is the mediator of the AB interaction. In this Letter, we propose a setup
in which the spectrum of the virtual photons drastically affects the AB effect,
by placing two parallel conducting plates. The cutoff frequency appears
in the electromagnetic wave modes
(see e.g. Ch.~8 of Ref.~\onlinecite{jackson99}) and plays exactly the role 
of the massive photons propagating in two dimension that satisfies
the Proca equation. 
We show that the interaction between
a charge and a magnetic flux is short-range due to the finite effective
mass of the photons,
leading to the short-range AB effect.

{\em Quantum electrodynamic Hamiltonian of the system-}.
The system under consideration is an AB interferometer of a charge ($e$)
moving around a magnetic flux
($\Phi$), where the motion of the charged particle is confined 
between two ideal conducting plates separated by a distance $d$~(see Fig.~1). 
We adopt QED to describe the system, which implies that
the interaction between the two bodies is mediated by 
vacuum photons ($\gamma$).  The Hamiltonian of the system is given by
\begin{equation}
\label{eq:H}
  H = \frac{1}{2m} (\mathbf{p} - \frac{e}{c}\mathbf{A})^2
     - \frac{\Phi}{4\pi} \int \mathbf{B} \cdot d\mathbf{r}'
     + H_\mathrm{em} \,,
\end{equation} 
where $\mathbf{A}$ and $\mathbf{B}=\nabla\times\mathbf{A}$ are the
vector potential and magnetic field of the radiation, respectively.
$H_\mathrm{em}$ represents the radiation modes. 
Note that there is no direct interaction between $e$ and $\Phi$.
In our system the radiation field is confined between
the conducting plates. 
The spatial part (vector potential) of the radiation
field,  
\begin{subequations}
\label{eq:A}
\begin{equation}
  \mathbf{A}(\mathbf{x},t) =
  \sum_{\mathbf{k},\lambda} \alpha_\mathbf{k}
  \left[
    u_{\mathbf{k}\lambda} (\mathbf{x},t) a_{\mathbf{k}\lambda} 
   +u_{\mathbf{k}\lambda}^*(\mathbf{x},t) a_{\mathbf{k}\lambda}^\dagger 
  \right] \vec{e}_\lambda \,,
\end{equation}
with the coefficient $\alpha_\mathbf{k}=\sqrt{2\pi\hbar c^2/\omega}$, 
is expanded by the eigenfunctions for the wave vector $\mathbf{k}$, 
$u_{\mathbf{k}\lambda} (\mathbf{x},t) 
 = u_{\mathbf{k}\lambda} (\mathbf{x}) e^{-i\omega t}$,
where
\begin{equation}
\label{eq:u}
 u_{\mathbf{k}\lambda} (\mathbf{x}) 
   = \left\{ \begin{array}{ll}
              \sqrt{\frac{2}{V}} e^{i\vec{\kappa}\cdot\vec{\rho}} 
                \sin{k_nz} \,, & \lambda=1,2 \\
              \sqrt{\frac{2}{V}} e^{i\vec{\kappa}\cdot\vec{\rho}} 
                \cos{k_nz} \,. & \lambda=3 
             \end{array}
     \right..
\end{equation}
\end{subequations}
For each mode $n$ ($=0,1,2,\cdots$) 
the eigenfrequency $\omega = ck = c\sqrt{\kappa^2+k_n^2}$
has the cutoff value of $ck_n$, where $k_n=n\pi/d$ is the quantized value
along the vertical direction of $\mathbf{k}$.
The Lorenz gauge is used here. This means that the radiation field has three 
polarization components, $\vec{e}_\lambda$ ($\lambda=1,2,3$). 
Eq.~\eqref{eq:A} is derived from the boundary conditions 
of the Maxwell equations 
for $\mathbf{A}$, where $\vec{\rho}$ and $\vec{\kappa}$ are the 
planar components
of $\mathbf{x}$ and $\mathbf{k}$ respectively.

%
%
%
Each mode of the radiation field satisfies the wave equation, 
\begin{equation}
\label{eq:wave}
 \left( \frac{\partial^2}{c^2\partial t^2} - \nabla_\|^2  \right)
    u_{\mathbf{k}\lambda} (\mathbf{x},t) 
  + k_n^2 u_{\mathbf{k}\lambda} (\mathbf{x},t) 
  = 0 \,,
\end{equation}
where $\nabla_\|^2 = \partial^2/\partial x^2 + \partial^2/\partial y^2$
is the 2-dimensional Laplacian in the plane parallel to the conducting
plates.
Note that Eq~\eqref{eq:wave} for $n\ne0$ is equivalent to 
the Proca equation describing massive photons~(see e.g., ch.2 of 
Ref.~\onlinecite{ryder96}) confined in two dimension,
where the mass is replaced by $\hbar k_n/c$. Therefore, the properties of 
the massive photons come into play in this system, {\em e.g.}, 
the group velocity
of the wave packet is less than $c$, etc. A particularly interesting
property is the short-range electrodynamic interaction due to
the finite mass of the spectrum. A photon mode with an angular frequency
lower than
$ck_n$ cannot propagate longer than the distance $\sim k_n^{-1}$. 
The two-body interaction is mediated by the exchange of virtual
photons, and the mass gap makes the interaction short-range, analogous
to the Yukawa interaction.

There is a gapless TM mode ($n=0$ and $\lambda=3$ in $u_{\mathbf{k}\lambda}$), 
but this mode is not coupled to the magnetic field of the flux perpendicular to
the conducting planes and therefore does not affect the nature of the
interaction between the two bodies.
The details of the short-range interaction are described below. 

{\em Derivation of the effective interaction-}. 
The Hamiltonian (Eq.~\eqref{eq:H}) can be rewritten as
\begin{subequations}
 \label{eq:H01}
\begin{equation}
 H = H_0 + H_1 ,
\end{equation}
where
\begin{equation}
 H_0 = \frac{\mathbf{p}^2}{2m} + H_\mathrm{em} 
\end{equation}
is the noninteracting part,
and the interaction consists of two parts:
\begin{equation}
 H_1 = V_a + V_b ,
\end{equation}
where
\begin{equation}
 V_a = -\frac{e}{mc} \mathbf{A}\cdot\mathbf{p}
\end{equation}
and
\begin{equation}
 V_b = - \frac{\Phi}{4\pi} \int \mathbf{B}\cdot d\mathbf{r}'
  \label{eq:V_b}
\end{equation}
\end{subequations}
represent the charge-vacuum and flux-vacuum interactions, respectively.
In the charge-potential interaction ($V_a$), we have omitted
the term $(i\hbar/2mc) \nabla\cdot\mathbf{A} + (e^2/2mc^2)\mathbf{A}^2$ which
is independent of the charge variable and irrelevant to the present
study.

To derive the effective interaction between the two bodies,
we use the canonical transformation technique 
(see also Ref.~\onlinecite{kang22}).
In the transformed Hamiltonian
$\tilde{H} = e^{-S} H e^{S}$, the first-order interaction 
is eliminated by imposing the condition
\begin{equation}
 H_1 + [H_0, S] = 0 ,
\end{equation}
from which we obtain (up to 2nd order)
\begin{subequations}
\label{eq:H_eff}
\begin{eqnarray}
 \tilde{H} &=& H_0 + H_2 \,, \\
  H_2 &=& \sum_\gamma |0\rangle
   \frac{ \langle0|V_a|\gamma\rangle\langle\gamma|V_b|0\rangle
         + \mathrm{h.c.} }{ -\hbar\omega}  \langle0| \,,
 \label{eq:H2}
\end{eqnarray}
\end{subequations}
where $|0\rangle$ denotes the radiation vacuum and $|\gamma\rangle =
a_{\mathbf{k}\lambda}^\dagger|0\rangle$.
Self-interaction terms are irrelevant and are therefore neglected in the
effective interaction $H_2$. 

Let us evaluate $H_2$ for the geometry of Fig.~1.   
The position of the charge is given in cylindrical 
coordinates, $\mathbf{r} = \rho\hat{\rho} + \rho\theta\hat{\theta}+z\hat{z}$.
The magnetic flux is along the $z$-axis, 
$\mathbf{r}'=z'\hat{z}$. 
The matrix elements $\langle0|V_a|\gamma\rangle$ and 
$\langle\gamma|V_b|0\rangle$
can be evaluated using the mode expansion of $\mathbf{A}$ and $\mathbf{B}$, 
and we find
\begin{equation}
 H_2 = \frac{ie\Phi}{2mc} F(\mathbf{r}) + \mathrm{h.c.} \,,
\end{equation}
where
\begin{equation}
 F(\mathbf{r}) = \sum_\mathbf{k} \frac{1}{k^2} \int_0^d dz'
    u_{\mathbf{k}1}(\mathbf{r}) u_{\mathbf{k}1}^*(\mathbf{r}') 
    \kappa \left(\mathbf{p}\cdot\hat{\theta} \cos{\phi} 
          - \mathbf{p}\cdot\hat{\rho} \sin{\phi} \right) \,.
\label{eq:F}
\end{equation}
Here $\phi$ is the angle between the vectors $\vec{\kappa}$ 
and $\mathbf{r}$. The second term of the integral in Eq.~\eqref{eq:F} vanishes 
and we get a simple form,
\begin{equation}
 H_2 = -\frac{e}{mc} \mathbf{p}\cdot\mathbf{a} \,,
\end{equation}
where 
\begin{equation}
  \mathbf{a} = \frac{2\Phi}{\pi d} 
   \sum_{n=\mathrm{odd}} \sin{(k_n z)} K_1(k_n\rho) \hat{\theta} 
\label{eq:a}
\end{equation}
plays the role of an effective vector potential.

As in the standard AB effect, the charged particle gets a phase shift 
\begin{equation}
 \phi_\mathrm{AB} = \frac{e}{\hbar c} \oint\mathbf{a}\cdot d\mathbf{r} \,,
\end{equation}
for a path moving around the magnetic flux.
For a circular path with radius $R$, we can get a simple expression:
\begin{equation}
\label{eq:phi-AB}
 \phi_\mathrm{AB} = \phi_\mathrm{AB}^0 
   \frac{4R}{d} \sum_{n=\mathrm{odd}} \sin{(k_n z)} K_1(k_nR)  \,,
\end{equation}
where $\phi_\mathrm{AB}^0=\frac{e\Phi}{\hbar c}$ 
is the usual AB phase in free space. 
Fig.~2 shows the AB phase as a function of distance $R$. Obviously,
the phase reduces to $\phi_\mathrm{AB}^0$ for $R\ll d$.
The effective mass (cutoff frequency) of the vacuum radiation modes 
comes into play and $\phi_\mathrm{AB}$ decreases monotonically 
with increasing $R$.
In the limit of $R\gg d$ it decays exponentially as
\begin{equation}
 \phi_\mathrm{AB} \sim \phi_\mathrm{AB}^0  \sqrt{\frac{8R}{d}}
  \sin{(k_1 z)} \exp{(-\pi R/d)}  \,,
\end{equation}
resulting from the nonzero mass of the mediating photons. 
Each mode of virtual photons has a different effective mass 
$\hbar k_n/c$, as shown in Eq.~\eqref{eq:wave},
but the behavior of $\phi_\mathrm{AB}$ at large $R/d$ is determined by the
contribution of the lowest mode ($n=1$).

{\em Discussion-}.
The reduced AB effect obtained here is a
manifestation of the modified photon spectrum. 
This demonstrates the role of the vacuum field in mediating
the interaction between charge and flux. It should be noted, however, 
that this is not the only way to describe the system. 
Moreover, the validity of our result can be questioned on the basis of the
topological property of the AB effect: the particle with charge
$e$ loops around the flux $\Phi$, and this should give the phase shift
of $e\Phi/\hbar c$ independent of the path. This inconsistency should be 
resolved.  

It is also possible to describe the system using a semiclassical approach. 
In the semiclassical approach, only the charged particle is
treated quantum mechanically, the vacuum photons are not included,
and the interaction between
$e$ and $\Phi$ is modified by the charges induced on the
conducting planes (not by changing the spectrum of the 
mediating photons, see Fig.~3).
The semiclassical Lagrangian of the system is given by 
$L_0 + L_\mathrm{int}$, where
$L_0$ is the kinetic part, and the interaction term
\begin{equation}
 L_\mathrm{int} = \frac{e}{c} \dot{\mathbf{r}} \cdot \mathbf{A}
  + \frac{1}{c}\int \sigma_\mathrm{in}(\tilde{\mathbf{r}}') \,
     \dot{\mathbf{r}}'\cdot\mathbf{A}(\mathbf{r}') d^2\tilde{\mathbf{r}}'  
\end{equation} 
consists of two parts. The charged particle ($e$) and 
the induced charge distribution on the conducting plates 
(denoted by $\sigma_\mathrm{in}$) interact with the flux ($\Phi$)
via the semiclassical vector potential. 
The variable $\tilde{\mathbf{r}}' \equiv \mathbf{r}'-\vec{\rho}$ 
denotes the position of the induced charges relative to $e$. 
The vector potential 
$\mathbf{A} = \hat{\theta} \Phi/2\pi\rho$ generated by the flux is not 
changed by the conducting plates, in contrast to the effective vector
potential (Eq.~\eqref{eq:a}) derived in the QED approach. 

As the charged particle moves along the path in the interferometer,
the induced charges on the conducting plates follow its motion
and also contribute to the phase shift. 
The density of the induced charges can be calculated from classical 
electrodynamics, and we find
\begin{equation}
 \sigma_\mathrm{in}(s) = -\frac{2e}{d^2} \sum_{n=\mathrm{odd}}
    n\sin{(k_nz)} K_0(k_n s) \,,
\end{equation} 
where $s$ is the planar distance from the position of the charged
particle $e$.  For a circular
orbit of the particle with radius $R$, the induced charges satisfying 
$s<R$ form a closed loop around the flux and contribute to the AB phase,
resulting in an effective charge of 
$e^*= 2\pi\int_0^R s\sigma_\mathrm{in}(s) ds$. We find
\begin{equation}
 e^* = -e + \frac{4\pi R}{d} \sum_{n=\mathrm{odd}} \sin{(k_nz)} K_1(k_n R) \,,
\end{equation}
and the net AB phase shift of the system is given by
\begin{equation}
 \phi_\mathrm{AB} = \frac{1}{\hbar c} (e+e^*)\Phi .
\end{equation} 
This leads to the result (Eq.~\eqref{eq:phi-AB}) which is identical 
to the one obtained from the QED approach.

The equivalence of $\phi_\mathrm{AB}$ derived from the two different approaches 
raises an intriguing question about the reality of virtual photons.
With QED we have shown that the exponential reduction of the AB phase
results from the modified photon spectrum. The confinement
of the vacuum between two ideal conducting plates provides the finite
effective mass of the mediating photons, and this leads to the short-range
interaction between the two bodies. In short, the mediating photons are
essential to understanding the nature of the two-body interaction. 
In contrast,
the reduction of the AB phase described in the semiclassical approach is
explained by the charges induced 
on the conducting plates, which also contribute to the phase shift of the
interference fringe. 

A similar situation is found in the Coulomb interaction between two
charges modified by conducting plates (Fig.~4). In the
QED approach, the Coulomb interaction is formed by the exchange of the
scalar component (denoted by $V$) of the vacuum radiation field. Each charge
interacts with $V$ and is described by the Hamiltonian
\begin{equation}
\label{eq:H_scalar}
 H = H_0 + e_1 V(\mathbf{r}_1) + e_2 V(\mathbf{r}_2) + H_\mathrm{em} .
\end{equation}
In the same way as the spatial components ($\mathbf{A}$) 
given in Eq.~\eqref{eq:A},
the scalar potential is expanded by the eigenmodes with finite masses (cutoff
frequencies), 
\begin{equation}
  V(\mathbf{x},t) =
  \sum_{\mathbf{k}} \alpha_\mathbf{k}
  \left[
    u_{\mathbf{k}0} (\mathbf{x}) a_{\mathbf{k}0} e^{-i\omega t} 
   +u_{\mathbf{k}0}^*(\mathbf{x}) a_{\mathbf{k}0}^\dagger 
      e^{i\omega t} 
  \right] \,,
\end{equation}
where 
 $u_{\mathbf{k}0} (\mathbf{x}) = \sqrt{\frac{2}{V}} 
 e^{i\vec{\kappa}\cdot\vec{\rho}} \sin{k_nz}$, and 
$a_{\mathbf{k}0}$($a_{\mathbf{k}0}^\dagger$) is an operator that destroys 
(creates) a scalar
photon of the wave vector $\mathbf{k}$. 

Applying the canonical transformation~(Eq.~\eqref{eq:H_eff}) to the 
Hamiltonian (Eq.~\eqref{eq:H_scalar}), 
we get the 2nd order effective interaction 
between the two charges,
\begin{equation}
\label{eq:H2-scalar}
 H_2 = \frac{4e_1e_2}{d} \sum_{n=1}^\infty \left[
       \sin{(k_n z_1)}\sin{(k_n z_2)} K_0(k_n\rho) \right] \,.
\end{equation}
The main feature of this effective interaction is the exponential decay,
$H_2 \sim \exp{(-\pi\rho/d)}$, at a large distance ($\rho/d\gg 1$), 
due to the finite mass of the
mediating photons. This interpretation is fascinating
in that it would imply the reality of ``scalar photons" which can never appear 
in the form of real particles (see also Ref.~\onlinecite{marletto23}).
However, it is also possible to obtain Eq.~\eqref{eq:H2-scalar} 
from a classical treatment.
(See. e.g., Prob.3.20 of Ref.~\onlinecite{jackson99}). 
In classical electrodynamics, 
the interaction between two charged particles is effectively shielded by 
the charges induced on the conducting plates.
Therefore, as in the case of the magnetic AB, the modified interaction
derived above does not conclusively support the ``reality" of 
virtual photons. 

Finally, we note that there have been theoretical studies of the AB effect as
a tool to provide information about photon mass.
The usual AB effect can be used to 
provide a limit on the photon mass in free space, since massive photons
lead to a finite range of the AB interaction~\cite{boulware89}.  
The experimental setup for this should be much larger than the scale of
the table-top apparatus. On the other hand, our study provides a way to observe
the role of massive photons in the AB effect at the table-top scale 
by effectively modifying the spectrum of photon modes.

{\em Conclusion-}.
We have proposed a realistic setup that leads to a reduced Aharonov-Bohm 
effect mediated by massive photons. The finite effective mass 
of virtual photons is created by confining the vacuum between 
two ideal conducting plates, where the electromagnetic waves satisfy 
the two-dimensional Proca equation. 
Consequently, the effective interaction between charge and flux becomes 
short-range, and so does the AB effect. 
While this result is a manifestation
of the modified spectrum of virtual photons, it is also possible to
describe the effect as a result of induced charges on the conducting
plates.  Apart from QED's merit of elimination of ``action at a distance",
it is also possible to interpret the reduced AB effect without mediating
photons.  So the question remains about the 
``reality" of virtual photons. It seems very unlikely to find a way to 
demonstrate the reality of virtual photons (although it should be indirect)  
in a real experiment that cannot be described without them. 
We note that the same problem exists in a modified Coulomb interaction
between two charges in a confined vacuum.



 \bibliography{references}

\newpage
\begin{figure}
\centering
\includegraphics[width=8cm]{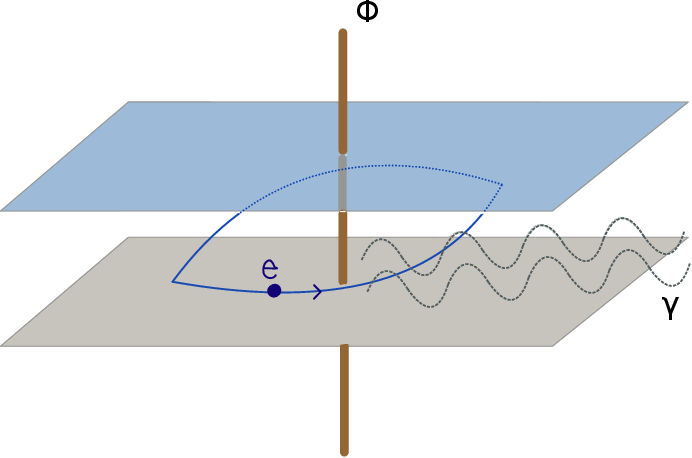} 
\caption{Schematic representation of the Aharonov-Bohm interference of
a charge $e$ confined between two ideal conducting plates. The interaction
between $e$ and the magnetic flux $\Phi$ is mediated by 
``massive" virtual photons ($\gamma$).
 }
\end{figure}

\begin{figure}
\centering
\includegraphics[width=8cm]{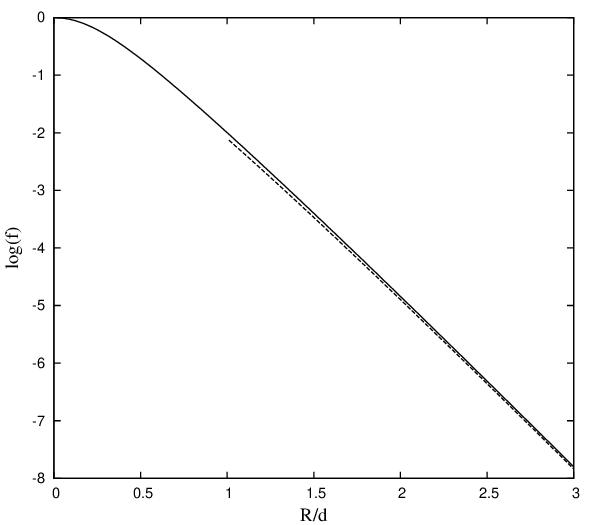} 
\caption{The dimensionless AB phase in the geometry of Fig~1, 
$f=\phi_\mathrm{AB}/\phi^0_\mathrm{AB}$
(in log scale), as a function of the radius of the loop (solid line).  
The reduced AB phase at a large distances is due to the
finite mass of the mediating photons. For comparison, the asymptotic 
behavior at large $R/d$, 
$f\simeq \sqrt{8R/d}\sin{(k_1z)}e^{-k_1R/d}$ (dashed line), is also shown.
}
\end{figure}

\begin{figure}
\centering
\includegraphics[width=8cm]{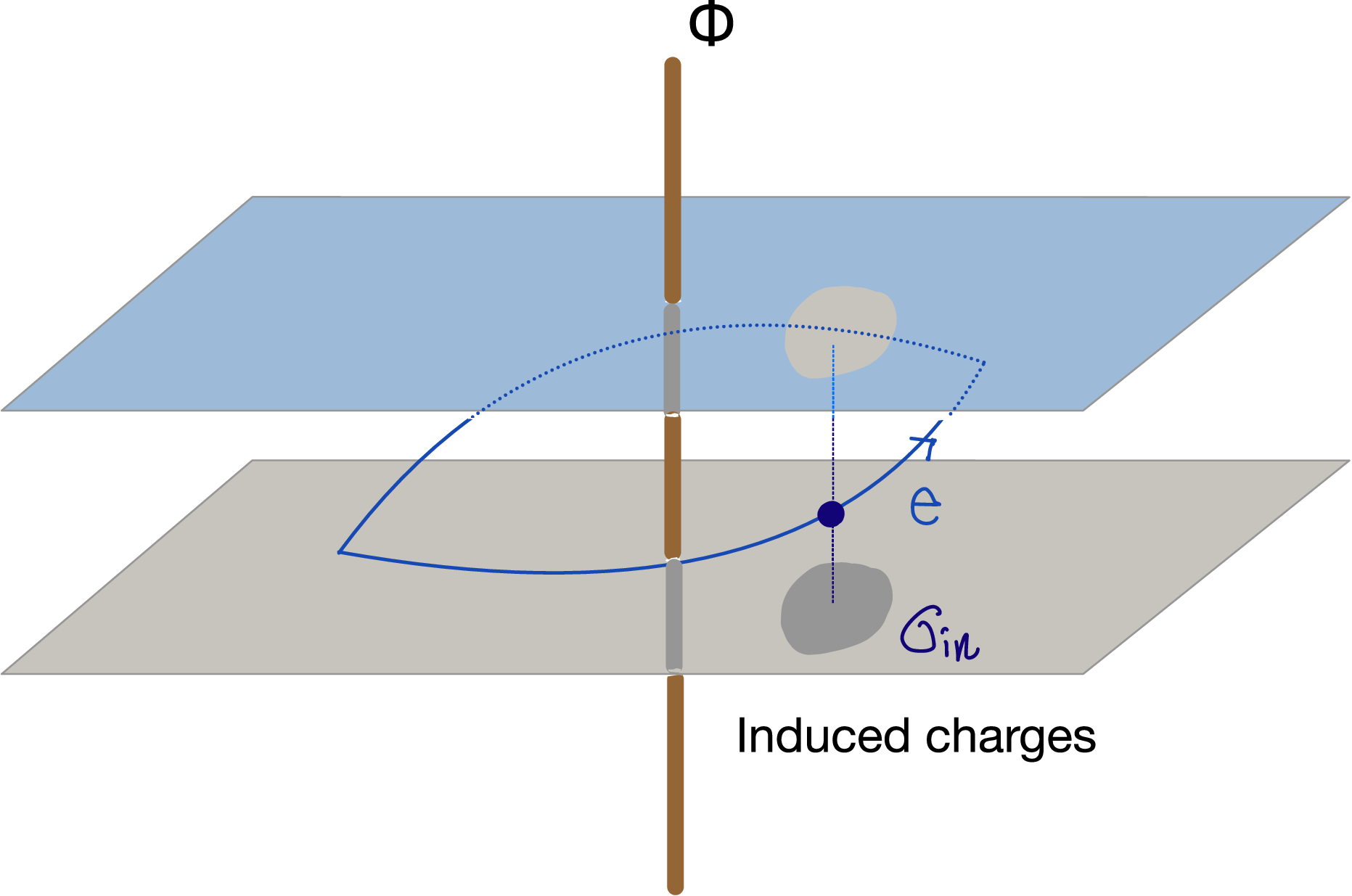}
\caption{Semiclassical picture of the reduced AB effect. The effective
interaction between $e$ and $\Phi$ is reduced by the charges induced 
on the conducting plates.
 }
\end{figure}

\begin{figure}
\centering
\includegraphics[width=8cm]{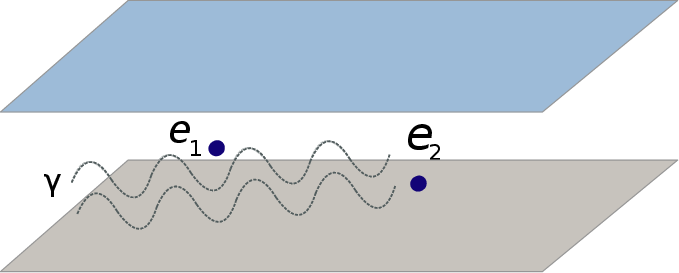}
\caption{Two point charges confined 
between conducting plates. The reduction of the Coulomb interaction 
between the  charges can be interpreted in two different ways: 
(1) the finite mass of 
the mediating scalar photons, or (2) the shielding by the induced charges on
the conducting plates.
 }
\end{figure}
\end{document}